# Exploring Avenues beyond Revised DSD Functionals: I. Range Separation, with *x*DSD as a Special Case

Golokesh Santra, Minsik Cho, and Jan M. L. Martin*

**ABSTRACT:** We have explored the use of range separation as a possible avenue for further improvement on our revDSD minimally empirical double hybrid functionals. Such ωDSD functionals encompass the XYG3 type of double hybrid (*i.e.*, *x*DSD) as a special case for ω → 0. As in our previous studies, the large and chemically diverse GMTKN55 benchmark suite was used for evaluation. Especially when using the D4 rather than D3BJ dispersion model, *x*DSD has a slight performance advantage in WTMAD2. As in previous studies, PBEP86 is the winning combination for the semilocal parts. *x*DSD$_n$-PBEP86-D4 marginally outperforms the previous "best in class" ωB97M(2) Berkeley double hybrid but without range separation and using fewer than half the number of empirical parameters. Range separation turns out to offer only marginal further improvements on GMTKN55 itself. While ωB97M(2) still yields better performance for small-molecule thermochemistry, this is compensated in WTMAD2 by the superior performance of the new functionals for conformer equilibria. Results for two external test sets with pronounced static correlation effects may indicate that range-separated double hybrids are more resilient to such effects.

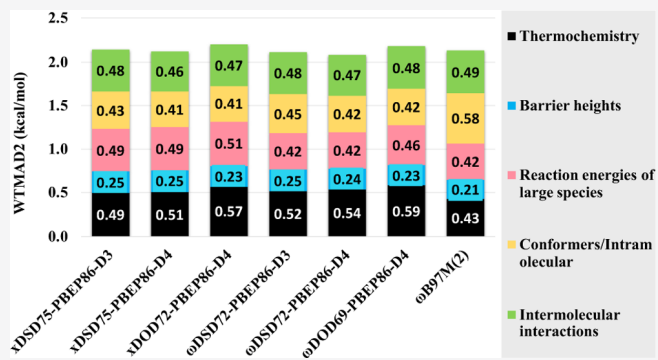

## 1. INTRODUCTION

Kohn–Sham density functional theory (KS-DFT)[1] is presently by far the most widely used family of electronic structure methods. Its combination of reasonable accuracy and comparatively gentle computational cost scaling makes it an appealing choice for medium and large molecules; for small molecules, wavefunction *ab initio* (WFT) approaches still outperform it.[2,3]

The accuracy of KS-DFT stands or falls with the exchange–correlation (XC) functional. Perdew[4] organized the plethora of available approaches into what he called a "Jacob's Ladder", arranged by the kinds of information employed in it: local density approximation (LDA) on the first rung, GGAs (generalized gradient approximations) on the second rung, meta-GGAs on the third rung (adding either the density Laplacian or the kinetic energy density), and hybrid functionals on the fourth rung (adding also the occupied orbital information). The fifth rung corresponds to the inclusion of virtual orbital information: the most widely used class of such methods are the so-called double hybrids (see refs [5–7] for reviews and most recently ref [8] by the present authors). As shown in refs,[8,9] their accuracy over the very large and diverse GMTKN55 (general main group thermochemistry, kinetics, and noncovalent interactions, with 55 problem sets) test suite[10] approaches that of WFT methods, yet the CPU cost increase over that of ordinary hybrid GGAs is actually quite modest if an RI (resolution of the identity[11,12]) approximation is applied in the MP2 (second-order Møller–Plesset) part.

Generally speaking, there are two basic approaches available for double hybrids in the literature, which we shall denote gDH (after Grimme[13]) and xDH (after the XYG3 functional[14,15]) in the article. In gDH, an iterative KS calculation is carried out with a fraction ($c'_{X,HF}$) of Hartree–Fock (HF) exchange and (1 − $c'_{X,HF}$) of DFA (density functional approximation) exchange, plus the DFA correlation scaled by a coefficient $c_{C,DFA}$. Next, using the converged orbitals from the KS step, a post-HF GLPT2 (second-order Görling–Levy perturbation theory)[16] correlation energy term is evaluated on the basis of the KS orbitals and added in. (As with lower-rung DFT methods, a dispersion correction can optionally be added, though it generally needs a prefactor that is less than unity, since some dispersion is already captured in the GLPT2 term.) Double hybrids with nonlocal correlation terms other than PT2, such as the *direct* random phase approximation (dRPA, see ref [17] and references therein), are discussed by Kállay and cow-orkers[18,19] as well as in the companion article[20] to the present work, while Manby and coworkers[21,22] very recently proposed a novel approach based on the Unsöld approximation (UW12).

In contrast, for xDHs, the KS orbitals used for the evaluation of all energy terms at the final step are evaluated for a standard hybrid with the full DFA correlation (i.e., $c_{C,DFA}$ = 100%) and with $c_X$ as appropriate for a typical hybrid functional. It was argued[14,15] that such orbitals are more appropriate as a basis for GLPT2 than the damped-correlation orbitals in the gDH, though this argument has been refuted on empirical grounds by Goerigk and Grimme[23] and by Kesharwani et al.[24]

Kozuch and Martin[25,26] modified the gDH approach into their dispersion-corrected spin-component-scaled double hybrids (DSDs), which employ the following energy equation

$$E_{DH-DSD} = E_{N1e} + c_{X,HF}E_{x,HF} + (1 - c_{X,HF})E_{X,XC}$$
$$+ c_{C,XC}E_{C,XC} + c_{2ab}E_{2ab} + c_{2ss}E_{2ss}$$
$$+ E_{disp}[s_6, s_8, c_{ATM}, a_1, a_2, etc.] \quad (1)$$

where $E_{N1e}$ stands for the sum of nuclear repulsion and one-electron energy terms; $c_{X,HF}$ and $c_{C,XC}$ are the fractions of exact exchange and semilocal correlation, respectively; $E_{disp}$ is the dispersion correction term (dependent upon parameters such as $s_6$, $s_8$, $a_1$, $a_2$, $c_{ATM}$, and so on); and $c_{2ab}$ and $c_{2ss}$ are the two coefficients corresponding to the opposite-spin and same-spin GLPT2 correlation, respectively. The xDH version thereof, denoted xDSD, has been explored in ref 24 and found to offer only a minor advantage over the corresponding DSD. It must however be said that both the DSD and the xDSD functionals were originally parameterized and validated using quite modest training sets (for reasons of computational cost); furthermore, the weighting of the subsets is somewhat arbitrary, and experimentation on our part showed considerable dependence of the final parameters on the weights used there. In contrast, the much larger GMTKN55 dataset is not only over 10x larger but uses a more robust, unambiguously weighted performance metric in the guise of WTMAD2 (weighted mean absolute deviation, type 2, in which the weights of the subsets are corrected for the different energy scales of the reference data). In ref 9, we were able to leverage GMTKN55 to obtain a family of more accurate revDSD functionals, with revDSD-PBEP86-D4 as the winner among them: just for the PBEP86 case, we also considered a single example of the xDH type and did find xrevDSD-PBEP86-D4 to be slightly more accurate still than revDSD-PBEP86-D4.

One objective of the present article is to explore whether this is true more generally: specifically, we shall investigate xDSD-PBEPBE here; we also include xDSD-PBEW91, xDSD-PBEB95, xDSD-BLYP, and xDSD-SCAN in our arsenal. Two types of dispersion correction, D3(BJ)[27,28] and more recent, flexible, and accurate D4,[29,30] will be considered, the latter with different many-body dispersion terms also. (We will also consider xDOD forms, in which same-spin GLPT2 is eliminated: this permits further acceleration for large systems through a Laplace transform algorithm.[31−36])

The second objective is to investigate whether revDSD can be improved through introducing the range-separated HF exchange (RSH). In the long-distance limit, the exchange potential of global hybrids (GHs) behaves[37] like $-c_X/r_{12}$ rather than the correct $-1/r_{12}$ term ($r_{12}$ being the interelectronic distance). Hence, Hirao and coworkers[38] proposed a scheme where the interelectronic repulsion operator $1/r_{12}$ is partitioned into a short-range (SR) component to be treated by a (meta)GGA and a long-range (LR) component to be treated by the "exact" exchange, and to "cross-fade" from the SR to LR component using an error function (erf $x$) and the complementary error function (erfc $x$ = 1 − erf $x$) of $r_{12}$. A more generalized form of this model was later proposed by Handy and coworkers[37]

$$\frac{1}{r_{12}} = \underbrace{\frac{1 - \alpha - \beta \text{erf}(\omega r_{12})}{r_{12}}}_{SR=\text{short-range}} + \underbrace{\frac{\alpha + \beta \text{erf}(\omega r_{12})}{r_{12}}}_{LR=\text{long-range}} \quad (2)$$

In this equation, $\omega$ represents the range separation parameter, which controls the transition between the LR and SR parts, $\alpha$ is the percentage of "exact" HF exchange in the short-range limit, and $\alpha + \beta$ is the corresponding percentage in the long-range limit. (Proper asymptotic behavior can be enforced through[39,40] $\alpha + \beta = 1/\varepsilon_0$, where the dielectric constant $\varepsilon_0 = 1$ in vacuo—leading to $\beta = 1 - \alpha$—and $\varepsilon_0 \to \infty$ for a perfect conductor.) $\omega$ can be determined empirically using a training set[37,41−45] or tuned non-empirically by minimizing the deviation from the conditions the exact KS functional must obey.[46,47] Following this approach, several empirical and non-empirical LC-DH functionals have been proposed, such as LC-PBE,[43] LC-ωPBE,[48] M11,[44] CAM-B3LYP,[37] ωB97,[45] ωB97X,[45] ωB97X-V,[49] ωB97M-V,[50] and many more.

We shall denote DSD-type double hybrids with RSH functionals ωDSD, where ω stands for the range separation parameter. Note that for $\omega = 0$, ωDSD and ωDOD functionals reduce to the xDSD and xDOD forms, respectively, which ties our two objectives together.

The combination of range separation with GLPT2 for the correlation energy was first proposed by Ángyán and coworkers.[51] Chai and Head-Gordon[52] instead obtained orbitals from an RSH calculation and then evaluated the GLPT2 correlation on the basis of these orbitals, the final energy being a mix of the GGA exchange, HF exchange, GGA correlation, and GLPT2 correlation. Their most recent elaboration of this concept was the ωB97M(2) functional,[53] which for the GMTKN55 benchmark was found to have the lowest WTMAD2 of all functionals surveyed.[8] (To be fair, however, it has three times the number of empirical parameters of the next best performer, xrevDSD-PBEP86-D4.[8])

Another effort along these lines was RSX-QIDH by Adamo et al.,[54] who established a "nonempirical" parameterization combining their quadratic integrand double hybrid (QIDH)[55] model with Savin's[56] RSX (range-separated exchange) scheme. Later, they introduced another such LC-DH, RSX-0DH.[57] Two very recent empirical RSDHs, originally developed for an electronic excitation energy benchmark, are ωB2PLYP and ωB2GP-PLYP by Goerigk and coworkers.[58]

## 2. COMPUTATIONAL METHODS

**2.1. Reference Data.** We can divide the parameter space into "linear" parameters such as $s_6$, $s_8$, $c_{2ss}$, and $c_{2ab}$ and "nonlinear" parameters such as $\alpha$ and $\omega$: every change in the latter requires complete recalculation of the entire GMTKN55 database, which would make a complete survey of the $(\alpha, \omega)$ parameter space for every underlying semilocal functional intractably costly. Fortunately, Gould[59] obtained so-called "diet" versions of GMTKN55, which are statistical reductions of the most representative 50 (diet50), 100 (diet100), or 150 (diet150) reactions.

After some experimentation, we settled on diet100 for the prescreening stage: based on this, we will decide which semilocal functionals to retain for in-depth investigation with the full GMTKN55 set.

GMTKN55[10] is the updated and larger form of the Grimme group's previous GMTKN24[60] and GMTKN30[23] databases. This dataset consists of 55 types of chemical problems, which can be further categorized into five top-level subsets: thermochemistry of small- and medium-sized molecules, barrier heights, large molecule reactions, intermolecular interactions, and conformer energies. One full evaluation of the GMTKN55 needs 2459 single-point energy calculations (give or take a few duplicates) to generate 1499 unique energy differences. (Complete details of all 55 subsets and original references can be found in Table S1 in the Supporting Information.)

Originally proposed by Goerigk et al.,[10] WTMAD2 has been used as the primary metric for this work:

$$\text{WTMAD2} = \frac{1}{\sum_{i=1}^{55} N_i} \cdot \sum_{i=1}^{55} N_i \cdot \frac{56.84 \text{ kcal/mol}}{|\overline{\Delta E}|_i} \cdot \text{MAD}_i \quad (3)$$

where $|\overline{\Delta E}|_i$ is the mean absolute value of all the reference energies from $i = 1$ to 55, $N_i$ is the number of systems in each subset, and $\text{MAD}_i$ is the mean absolute difference between calculated and reference energies for each of the 55 subsets. Note that, from the statistical viewpoint, MAD (mean absolute deviation) is a more "robust" metric than rmsd (root-mean-square deviation)[61] as MAD is more resilient to a small number of large outliers than rmsd. For a normal distribution without a systematic error, rmsd ≈ 5MAD/4.[62]

As one reviewer pointed out, the average absolute reaction energies for NBPRC and MB16-43 provided in the original GMTKN55 article[11] differ from the corresponding values calculated from the individual data supplied in the Supporting Information. If these corrected average absolute reaction energies were employed in the construction of eq 3, then their average, which appears in eq 3 as the overall scale factor, would be 57.76 rather than 56.84. However, as all previously published articles on GMTKN55 (such as refs [8,9,28,63–66]) have used the original (56.84) coefficient, we are also retaining it for the sake of compatibility. It goes without saying that this will not affect the ranking between functionals; those who prefer WTMAD2$_{57.76}$ can simply multiply all WTMAD2 values by 1.0162.

Reference geometries were taken "as is" from ref 10 and not optimized further.

**2.2. Electronic Structure Calculations.** All the calculations were performed using the Q-CHEM 5.3[67] package (except ωB2GP-PLYP[58] and ωB2PLYP,[58] for which ORCA 4.2.1[68] has been used), running on the ChemFarm HPC cluster of the Weizmann Institute Faculty of Chemistry.

The Weigend–Ahlrichs[69] def2-QZVPP basis set was considered throughout with a few exceptions, such as the WATER27, RG18, IL16, G21EA, AHB21, BH76, and BH76RC subsets, where the diffuse-function-augmented def2-QZVPPD[70] basis set was used instead. However, for the computationally demanding C60ISO and UPU23 subsets, which have small weights in WTMAD2, the more economical def2-TZVPP[69] basis set was employed to curb the computational cost. The SG-3[71] integration grid was used across the board, except for the SCAN (strongly constrained and appropriately normed[72] meta-GGA type) variants, where due to SCAN's severe integration grid sensitivity,[73] an unpruned (150, 590) grid was employed. In the MP2-like step, the RI approximation was applied in conjunction with the def2-QZVPPD-RI fitting basis set.[74,75] For the ωB2GP-PLYP[58] and ωB2PLYP[58] functionals using ORCA, we have used the JK auxiliary basis set for Coulomb and exchange RI integrals (def2/JK).[76]

In this project, while most of the calculations were completed using frozen inner-shell orbitals, we made two departures from this recipe to avoid unacceptably small orbital energy gaps between the highest frozen and lowest correlated orbitals. First, for the MB16-43, HEAVY28, HEAVYSB11, ALK8, CHB6, and ALKBDE10 subsets, we correlated the $(n − 1)$sp subvalence electrons of the metal and metalloid atoms. Second, for HAL59 and HEAVY28, the $(n − 1)$spd orbitals of the heavy p-block elements were kept unfrozen. Note that unlike the valence correlation consistent basis sets, the Weigend–Ahlrichs QZVPP basis set is multiple-zeta in the core as well and contains core–valence polarization functions (see Table 1 of ref 69). Semidalas and Martin (in the context of composite wavefunction calculations) considered[3] the impact of the core–valence correlation on GMTKN55 using correlation-consistent core–valence basis sets and found that its impact is on the order of 0.05 kcal/mol—which will be further reduced here as the PT2 correlation terms are scaled down in a double hybrid.

**2.3. Optimization of Parameters.** Range-separated DSD double hybrids have seven empirical parameters:

a. Fraction of exact exchange $c_{X,HF}$ or $\alpha$.
b. Fraction of the semilocal DFT correlation $c_{C,DFT}$.
c. Fraction of the opposite-spin PT2 correlation $c_{2ab}$.
d. Fraction of the same-spin PT2 correlation $c_{2ss} = c_{2aa+bb}$.
e. Prefactor $s_6$ for the D3(BJ) dispersion correction.[27,28,77]
f. Damping function range parameter $a_2$ for D3(BJ) (as recommended in refs 26 and 78, we set $a_1 = 0$ and $s_8 = 0$).
g. The range separation parameter, $\omega$.

Now, the xDSD family of functionals, being the special case of the range-separated DSD type (i.e., $\omega = 0$), have six parameters (a–f) instead of seven.

Powell's BOBYQA (bound optimization BY quadratic approximation) derivative-free constrained optimizer[79] and a few scripts and Fortran programs developed in-house were used for optimizing the parameters.

Once a full GMTKN55 evaluation is finished with a fixed set of $\{c_{X,HF}, c_{C,DFT}, \omega\}$, no further electronic structure calculation is needed to get an associated optimal set of (c–f); the latter set of parameters can be obtained in what amounts to an inner optimization loop, whereas $c_{X,HF}$, $c_{X,DFT}$, and $\omega$ (where applicable) can be minimized in an outer optimization loop. (We previously found in the revDSD article[9] that the coupling between (a) and (c,d) is too strong to permit placing $c_{C,DFT}$ in the inner loop and that for a fixed value of (a) convergence of the DFT correlation parameter, up to two decimal places can be achieved within two *macroiterations*.) The process is analogous to *microiterations* versus *macroiterations* in CASSCF algorithms (CI coefficients vs orbitals, see ref 80 and references therein) or QM–MM geometry optimizations, where geometric parameters in the MM layer are subjected to microiterations for each change of the coordinates in the QM layer, and the latter are optimized in macroiteration cycles (e.g., ref 81).

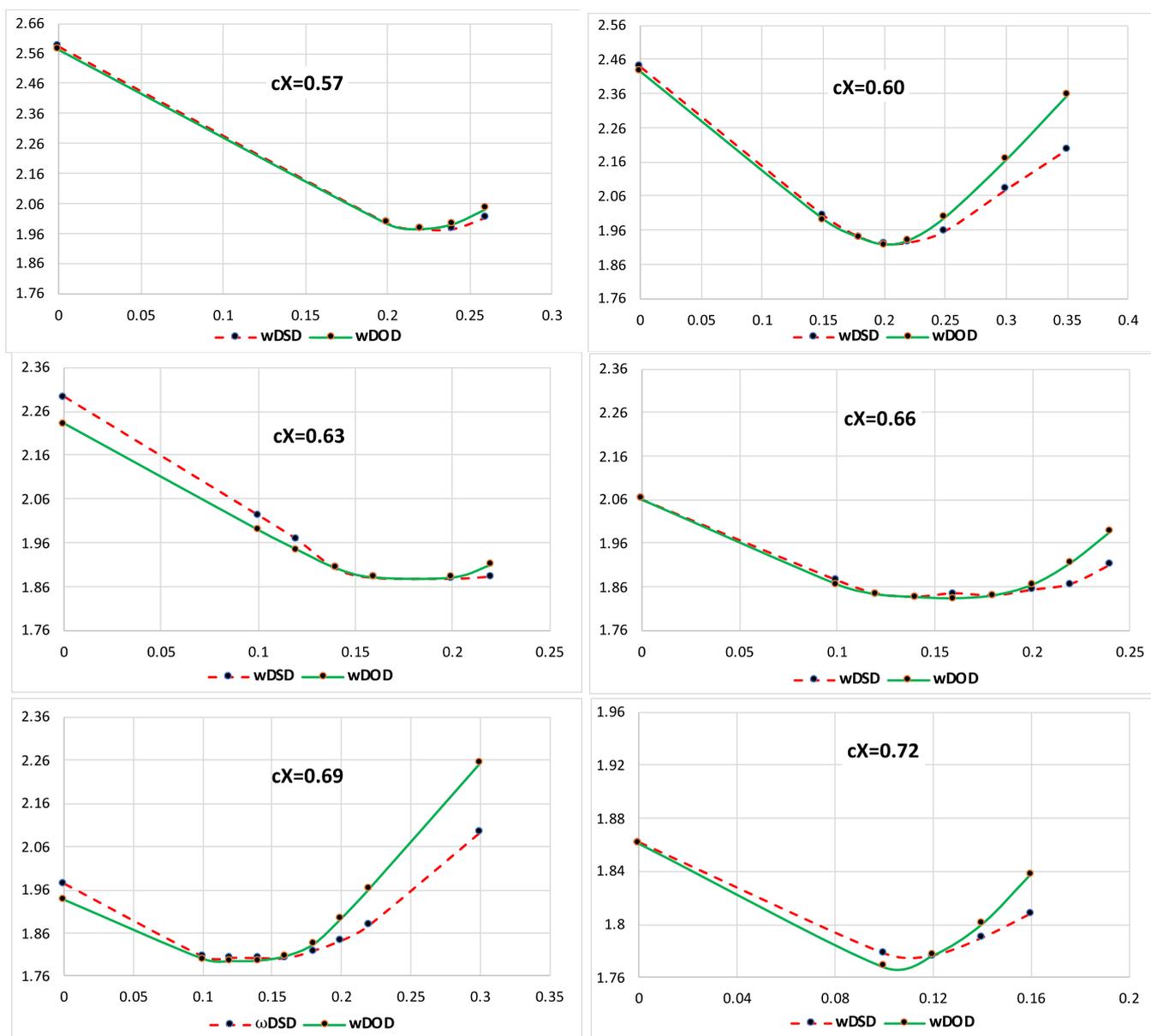

**Figure 1.** Change of WTMAD2 (kcal/mol) for $\omega DSD_X$ and $\omega DOD_X$-PBEP86-D3BJ with respect to range separation parameter $\omega$ (x-axis) for different $c_{X,HF}$ values. (Similar graphs for $\omega DSD_X$-PBEB95-D3BJ, $\omega DSD_X$-PBEPW91-D3BJ, and $\omega DSD_X$-PBEPBE-D3BJ and their $\omega DOD_X$ versions can be found in Figure S1 in the Supporting Information.)

The rate-determining step during the *microiterations* would normally be the evaluation of all the dispersion corrections for the entire GMTKN55 set, one after another, for a given combination of parameters. (When these were all done sequentially, the total wall clock time on our system was about 10−15 min for each microiteration, much of it due to the operating system overhead.) However, this step could be greatly accelerated by parceling out the individual D3BJ or D4 evaluations between all CPUs in a 40-core node. A minor I/O contention issue thus created was resolved by copying all required files onto a temporary RAM file system.

In the present case, the optimum value of the range separation parameter $\omega$ for a given fixed value of $\alpha$ is determined manually by interpolation. We repeated this process for six equally spaced $\alpha$ values, ranging from 0.57 to 0.72, to construct six different range-separated DSD (i.e., $\omega$DSD) functionals.

## 3. RESULTS AND DISCUSSION

### 3.1. Prescreening of Functionals with Diet-GMTKN55.

The prescreening experiment was performed using diet100 for $\omega DSD_n$-PBEP86-D3BJ, $\omega DSD_n$-PBEPBE-D3BJ, $\omega DSD_n$-PBEB95-D3BJ, and $\omega DSD_n$-PBEPW91-D3BJ variants, where n stands for the fraction of HF exchange used, that is, $c_{X,HF}$ or $\alpha$.

From Figures 1 and S1 in the Supporting Information, it is clear that, for every $c_{X,HF}$ considered, the $\omega DSD_n$ and $\omega DOD_n$-PBEP86-D3BJ variants benefited from range separation, whereas for the other exchange−correlation (XC) combinations, only $c_{X,HF}$ = 0.57 and to some extent $c_{X,HF}$ = 0.60 showed some advantage.

Table 1. WTMAD2 (kcal/mol) for $x\text{DSD}_n$ ($x\text{DOD}_n$)-XC-D3BJ (D4) Functionals and Final Parameters for the D4 Variants

| functionals | | WTMAD2 (kcal/mol) | | parameters | | | | | | | | |
|---|---|---|---|---|---|---|---|---|---|---|---|---|
| | | D3BJ[a] | D4 | $c_{X,HF}$ | $c_{C,DFT}$ | $c_{2ab}$ | $c_{2ss}$ | $s_6$ | $s_8$ | $c_{ATM}$ | $a_1$ | $a_2$ |
| $x$DSD | PBEP86 | 2.144 | 2.119 | 0.75 | 0.3517 | 0.6623 | 0.1168 | 0.4246 | [0] | 1.0 | 0.2828 | 4.7204 |
| | PBEB95 | 2.639 | 2.403 | 0.74 | 0.3887 | 0.6384 | 0.0470 | 0.4080 | [0] | 1.0 | 0.3247 | 3.8035 |
| | BLYP | 2.254 | 2.242 | 0.77 | 0.4527 | 0.6407 | 0.1025 | 0.5289 | [0] | 1.0 | 0.1357 | 5.0726 |
| | SCAN | 2.488 | 2.378 | 0.69 | 0.4431 | 0.6721 | 0.0128 | 0.3546 | [0] | 1.0 | 0.1913 | 5.0185 |
| | PBEPW91 | 2.373 | 2.203 | 0.72 | 0.4030 | 0.6738 | 0.0272 | 0.5929 | [0] | 1.0 | 0.3191 | 4.1913 |
| | PBEPBE | 2.420 | 2.238 | 0.72 | 0.4014 | 0.6798 | 0.0217 | 0.6091 | [0] | 1.0 | 0.3097 | 4.2792 |
| $x$DOD | PBEP86 | 2.231 | 2.196 | 0.72 | 0.3996 | 0.6490 | [0] | 0.5389 | [0] | 1.0 | 0.2094 | 5.0148 |
| | PBEB95 | 2.774 | 2.491 | 0.69 | 0.4422 | 0.6070 | [0] | 0.5155 | [0] | 1.0 | 0.3155 | 3.7246 |
| | BLYP | 2.564 | 2.543 | 0.74 | 0.5181 | 0.6925 | [0] | 0.7022 | [0] | 1.0 | 0.1080 | 5.0099 |
| | SCAN | 2.503 | 2.385 | 0.69 | 0.4541 | 0.6784 | [0] | 0.3709 | [0] | 1.0 | 0.0847 | 5.6326 |
| | PBEPW91 | 2.418 | 2.219 | 0.69 | 0.4459 | 0.6430 | [0] | 0.6739 | [0] | 1.0 | 0.2679 | 4.4855 |
| | PBEPBE | 2.451 | 2.243 | 0.69 | 0.4337 | 0.6462 | [0] | 0.6869 | [0] | 1.0 | 0.2713 | 4.4529 |

[a]All results are with fixed $a_2$ = 5.5, $a_1$ = 0, and $s_8$ = 0.

Now, when we repeated the same experiment for other XC combinations, we found that none reached the accuracy of $\omega\text{DSD}_n$ or $\omega\text{DOD}_n$-PBEP86-D3BJ in terms of WTMAD2 (see Table S3 in the Supporting Information). Therefore, we decided to proceed further with the range separation experiment only for PBEP86, where we considered the simultaneous variation of the short-range HF exchange ($c_X$) and range separation ($\omega$) parameters—using full GMTKN55.

**3.2. $x$DSD Functionals.** In our previous study,[9] for the xrevDSD-PBEP86-D4 functional, we did not reoptimize the $c_{X,HF}$ parameter and instead took the earlier reported best value by Martin and coworkers[24] and optimized other linear parameters against GMTKN55. What if we also vary $c_{X,HF}$? (Note that a full set of GMTKN55 electronic structure calculations is necessary for each $c_{X,HF}$ value considered.) In the current study, we seek the $c_{X,HF}$ that minimizes WTMAD2.

We denote these new functionals $x\text{DSD}_n$-PBEP86-Disp, where "$n$" stands for the percentage of HF exchange used. Following this notation, $x\text{DSD}_{69}$-PBEP86-D4 is the same as the xrevDSD-PBEP86-D4 functional reported by us in ref 9.

$x\text{DSD}_n$-PBEP86: We performed a full GMTKN55 evaluation for eight equally spaced $c_{X,HF}$ points, ranging from 0.57 to 0.78, followed by parameter optimization, taking both D3BJ and D4 dispersion corrections into account. Both with D3BJ and D4, $c_{X,HF}$ = 0.75 offered the lowest WTMAD2 instead of previously reported[24] $c_{X,HF}$ = 0.69 for $x$DSD-PBEP86—which could be an artifact of optimizing $c_{X,HF}$ against a training set considerably smaller than that of GNTKN55.

With the D3BJ dispersion correction, the WTMAD2 for $x\text{DSD}_{75}$-PBEP86-D3BJ is 2.144 kcal/mol—which is essentially identical to $\omega$B97M(2) (WTMAD2 = 2.131 kcal/mol) but with fewer empirical parameters and (still) no range separation. (The latter is significant, considering that many codes are able to exploit density fitting in GHs but not range-separated hybrids.) Here, we should note that the $\omega$B97M(2) functional was not trained against GMTKN55 but against a subset of the ca. 5000-point MGCDB84 (main group chemistry database[82]); however, a fair amount of overlap exists between GMTKN55 and MGCDB84. Increased percentages of HF exchange in our optimized functional (i.e., going from $c_{X,HF}$ = 0.69 for xrevDSD-PBEP86-D3BJ to $c_{X,HF}$ = 0.75 for $x\text{DSD}_{75}$-PBEP86-D3BJ) mainly benefited small-molecule thermochemistry and intermolecular interactions (see Table S4 in the Supporting Information). Now, when we constrained $c_{2ss}$ = 0 (i.e., the $x\text{DOD}_n$-PBEP86-D3BJ functionals), $c_{X,HF}$ = 0.72 offered the lowest WTMAD2 (=2.231 kcal/mol); small-molecule thermochemistry suffered most of the deterioration resulted from applying this constraint.

Next, we repeated the same experiment for five different semilocal XC combinations, namely, PBEPBE, PBEB95, PBEPW91, SCAN, and BLYP (see Table 1 for WTMAD2 statistics and optimized parameters). Although, as expected, all $x\text{DSD}_n$ functionals surpassed the corresponding revDSD variants, none approached the accuracy of $x\text{DSD}_{75}$-PBEP86-D3BJ; the only contender that came close was $x\text{DSD}_{77}$-BLYP-D3BJ with 77% HF exchange. For the $x\text{DSD}_n$-PBEPBE variants, one finds the "sweet spot" at $c_{X,HF}$ = 0.72, unlike the previously reported $c_{X,HF}$ = 0.68 in ref 24.

For the $x$DOD functionals (which permit us to use reduced-scaling PT2 algorithms[31−36] as well as eliminate one empirical parameter), all except $x\text{DOD}_{69}$-SCAN-D3BJ prefer a lesser percentage of exact exchange than the corresponding $x$DSD variants. The largest penalty for restricting $c_{2ss}$ = 0 is paid by $x\text{DOD}_{74}$-BLYP-D3BJ—the WTMAD2 value drops from 2.254 kcal/mol to 2.564 kcal/mol. Upon further inspection, of all the 55 subsets, W4-11, TAUT15, and BSR36 are the three most affected ones. Similar to our previous observation for DSD-SCAN functionals,[9] $x\text{DOD}_{69}$-SCAN-D3BJ sacrifices almost nothing when constraining $c_{2ss}$ to be zero.

If, instead of using the fixed value $a_2$ = 5.5, we optimize it together with the other "inner loop parameters" (i.e., $c_{DFT}$, $c_{2ab}$, $c_{2ss}$, and $s_6$), the WTMAD2 for the $x$DSD-type functionals remains more or less unchanged, whereas for $x\text{DOD}_{69}$-PBEB95-D3BJ and $x\text{DOD}_{74}$-BLYP-D3BJ, WTMAD2 values decrease by 0.068 and 0.058 kcal/mol, respectively.

In our prior work,[9] for technical reasons, we adopted $c_{ATM}$ = $s_6$, where $c_{ATM}$ is the prefactor for the Axilrod−Teller−Muto (ATM)[83,84] three-body correction term. If we allow $c_{ATM}$ as a variable, this somewhat reduces the WTMAD2 for $x\text{DSD}_n$-PBEPBE-D4, $x\text{DSD}_n$-PBEPW91-D4, and $x\text{DSD}n$-PBEB95-D4 and their $x$DOD variants. This leaves us with five adjustable dispersion parameters for the $x$DSD-D4 functionals. When we optimized all of them along with other parameters using BOBYQA, we noticed $s_8$ settling on values close to zero and $c_{ATM}$ on values close to one. Hence, if we constrain $s_8$ = 0 and $c_{ATM}$ = 1, the loss in accuracy is negligible, which permits eliminating two adjustable parameters. The only exceptions are SCAN variants, where although the $s_8$ value is not close to zero and $c_{ATM}$ is more than two for all cases, restricting $c_{ATM} = 1$ and $s_8 = 0$ does not appreciably degrade the total WTMAD2: for $x\text{DSD}_{69}$-SCAN-D4, it just increases from 2.351 to 2.379 kcal/mol when we impose these restrictions.

Table 2. WTMAD2 (kcal/mol) and Final Recommended Parameters for the $\omega\text{DSD}_n$ ($\omega\text{DOD}_n$)-PBEP86-D4 Functionals

| functional | WTMA2 (kcal/mol) | | parameters | | | | | | | | | |
|---|---|---|---|---|---|---|---|---|---|---|---|---|
| | D3BJ | D4 | $\omega$ | $c_{X,HF}$ | $c_{C,DFT}$ | $c_{2ab}$ | $c_{2ss}$ | $s_6$ | $s_8$ | $c_{ATM}$ | $a_1$ | $a_2$ |
| $\omega$DSD-PBEP86 | 2.108 | 2.083 | 0.13 | 0.72 | 0.3425 | 0.6904 | 0.1343 | 0.4685 | [0] | 1.0 | 0.1884 | 5.0101 |
| | 2.112 | 2.089 | 0.16 | 0.69 | 0.3607 | 0.6610 | 0.1232 | 0.5078 | [0] | 1.0 | 0.1545 | 5.1749 |
| | 2.129 | 2.116 | 0.18 | 0.66 | 0.3748 | 0.6265 | 0.1222 | 0.5456 | [0] | 1.0 | 0.2127 | 4.7954 |
| | 2.170 | 2.154 | 0.20 | 0.63 | 0.3907 | 0.5944 | 0.1146 | 0.5848 | [0] | 1.0 | 0.1907 | 4.9182 |
| | 2.229 | 2.202 | 0.22 | 0.60 | 0.4124 | 0.5519 | 0.1220 | 0.6164 | [0] | 1.0 | 0.1326 | 5.3251 |
| | 2.289 | 2.258 | 0.22 | 0.57 | 0.4257 | 0.5262 | 0.0944 | 0.6689 | [0] | 1.0 | 0.1763 | 4.9845 |
| $\omega$DOD-PBEP86 | 2.220 | 2.184 | 0.08 | 0.72 | 0.3817 | 0.7056 | [0] | 0.5405 | [0] | 1.0 | 0.1498 | 5.2264 |
| | 2.204 | 2.175 | 0.10 | 0.69 | 0.3993 | 0.6676 | [0] | 0.5797 | [0] | 1.0 | 0.1442 | 5.2814 |
| | 2.214 | 2.176 | 0.15 | 0.66 | 0.4030 | 0.6483 | [0] | 0.6179 | [0] | 1.0 | 0.1186 | 5.2836 |
| | 2.232 | 2.199 | 0.16 | 0.63 | 0.4207 | 0.6095 | [0] | 0.6603 | [0] | 1.0 | 0.1561 | 5.1364 |
| | 2.279 | 2.241 | 0.18 | 0.60 | 0.4320 | 0.5760 | [0] | 0.6969 | [0] | 1.0 | 0.1241 | 5.2353 |
| | 2.361 | 2.302 | 0.20 | 0.57 | 0.4410 | 0.5631 | [0] | 0.7081 | [0] | 1.0 | 0.0971 | 5.3320 |

Hence, going forward, we decided to freeze $s_8$ and $c_{ATM}$ throughout and optimize the remaining parameters (see Table 2). With D4 dispersion, the best performer of the $x$DSD family is again $x\text{DSD}_{75}$-PBEP86-D4 (WTMAD2 = 2.119 kcal/mol), which now marginally outperforms Mardirossian and Head-Gordon's $\omega$B97M(2)[53] (2.131 kcal/mol). (To be fair, such a small difference is really within the uncertainty of the reference values, as discussed in ref 9.) Among all the 55 subsets, switching from D3BJ to D4 improved the performance of BUT14DIOL, HAL59, and MCONF subsets quite a bit for $x\text{DSD}_{74}$- and $x\text{DOD}_{69}$-PBEPBE-D4; BSR36, BUT14DIOL, and MCONF got improved for $x\text{DSD}_{72}$- and $x\text{DOD}_{69}$-PBEPW91-D4. For the PBEB95 XC combination, three subsets, BUT14DIOL, PCONF21, and MCONF, benefited from replacing D3BJ with the D4 dispersion correction. Lastly, for $x\text{DOD}_{69}$-PBEB95-D4, AMINO20X4, BSR36, BUT14DIOL, MCONF, and PCONF21 subsets benefited the most. (In response to a reviewer query, we have evaluated the impact of the recent revision[85] of D4, which corresponds to version 3 of the standalone dftd4 program, and found the difference for WTMAD2 to be a negligible 0.005 kcal/mol even for PBE0-D4, where $s_6 = 1$, unlike for the double hybrids.)

Except for $x\text{DSD}_{75}$-PBEP86-D4 and $x\text{DSD}_{77}$-BLYP-D4, all other functionals prefer a very small fraction of the opposite-spin MP2-like correlation. This is why for these functionals, we sacrifice very little by constraining it to zero, i.e., shifting from $x$DSD-D4 to $x$DOD-D4.

We should also mention the accidental similarity of $x\text{DSD}_{75}$-PBEP86-D4 to Kállay and coworkers' dRPA75[19] regarding the preferred percentage of exact exchange.

Can the performance of $x\text{DSD}n$-XC-D4 functionals be improved further by replacing the default three-body ATM term by the many-body MBD correction of Tkatchenko[86] and scaling it with a prefactor (now called $c_{MBD}$ rather than $c_{ATM}$)? While we did find some improvement for one specific case, namely, $x\text{DSD}_{74}$-PBEB95-D4MBD (the WTMAD2 drops from 2.403 to 2.288 kcal/mol), it appears that the molecules being considered here still are not large enough for higher-order MBD corrections to become statistically noteworthy and that our answer is hence inconclusive. (See Table S2 of the Supporting Information for some of our data.)

In a recent study, Semidalas and Martin[3] have reported significant improvement for their composite methods by switching from the frozen core to the core−valence correlation and using the complete basis set (CBS) extrapolation from aug-ccpwCVTZ(-PP) and aug-cc-pwCVQZ(-PP) level calculations. Hence, we also checked whether further improvement of WTMAD2 statistics is possible by using a sufficiently large basis set and including the subvalence correlation in the MP2-like part. Extrapolating from the core−valence aug-ccpwCVTZ(-PP) and aug-cc-pwCVQZ(-PP) energies for $x\text{DSD}_{75}$-PBEP86-D3BJ using the $L^{-3}$ formula for opposite-spin and $L^{-5}$ for same-spin MP2-like correlation, proposed by Halkier et al.,[87] we found a change in WTMAD2 of up to three decimal places (0.00014 kcal/mol). We have therefore not explored this further for other double hybrids.

**3.3. Range Separation.** We revisit the range separation experiment now using the full GMTKN55 database. With the D3BJ dispersion correction, we found the lowest WTMAD2 (2.108 kcal/mol) for $\omega\text{DSD}_{72}$-PBEP86-D3BJ ($\omega = 0.13$), which is very close to what $\omega\text{DSD}_{69}$-PBEP86-D3BJ ($\omega = 0.16$) exhibited (2.112 kcal/mol). In general, the reduction of $c_{X,HF}$ entails an increase in $\omega$ in compensation.

Similar to the previous section, here also, we checked how much performance we sacrificed by switching from $\omega$DSD to $\omega$DOD. With WTMAD = 2.204 kcal/mol, $\omega\text{DOD}_{69}$-PBEP86-D3BJ ($\omega = 0.10$) appeared to be the best performer in this category. By and large, we gave up about 0.1 kcal/mol accuracy by the constraint $c_{2ss} = 0$; small-molecule thermochemistry is consistently the category the most affected by this restriction. Upon further inspection over all the 55 subsets, we found that BSR36 and TAUT15 are the main sources of this degradation. Both $\omega\text{DSD}_{72}$-PBEP86-D3BJ ($\omega = 0.13$) and $\omega\text{DOD}_{69}$-PBEP86-D3BJ ($\omega = 0.10$) only marginally outperform the corresponding $\omega = 0$ variants $x\text{DSD}_{75}$-PBEP86-D3BJ and $x\text{DOD}_{72}$-PBEP86-D3BJ, respectively. Next, when instead of freezing $a_2$ at 5.5, we optimized it together with other parameters, we found almost no change in WTMAD2 statistics, neither for $\omega$DSD nor for $\omega$DOD.

Aiming for further improvement, we considered replacing the D3BJ term by D4 energy components.[30] Similar to what was mentioned earlier in this article, we found that imposing $s_8 = 0$ and $c_{ATM} = 1$ caused only an insignificant increase of WTMAD2, although here optimal $c_{ATM}$ was somewhat further from unity. The lowest WTMAD2 we can get by shifting from D3BJ to D4 is 2.083 kcal/mol for $\omega\text{DSD}_{72}$-PBEP86-D4—at the cost of eight adjustable parameters.

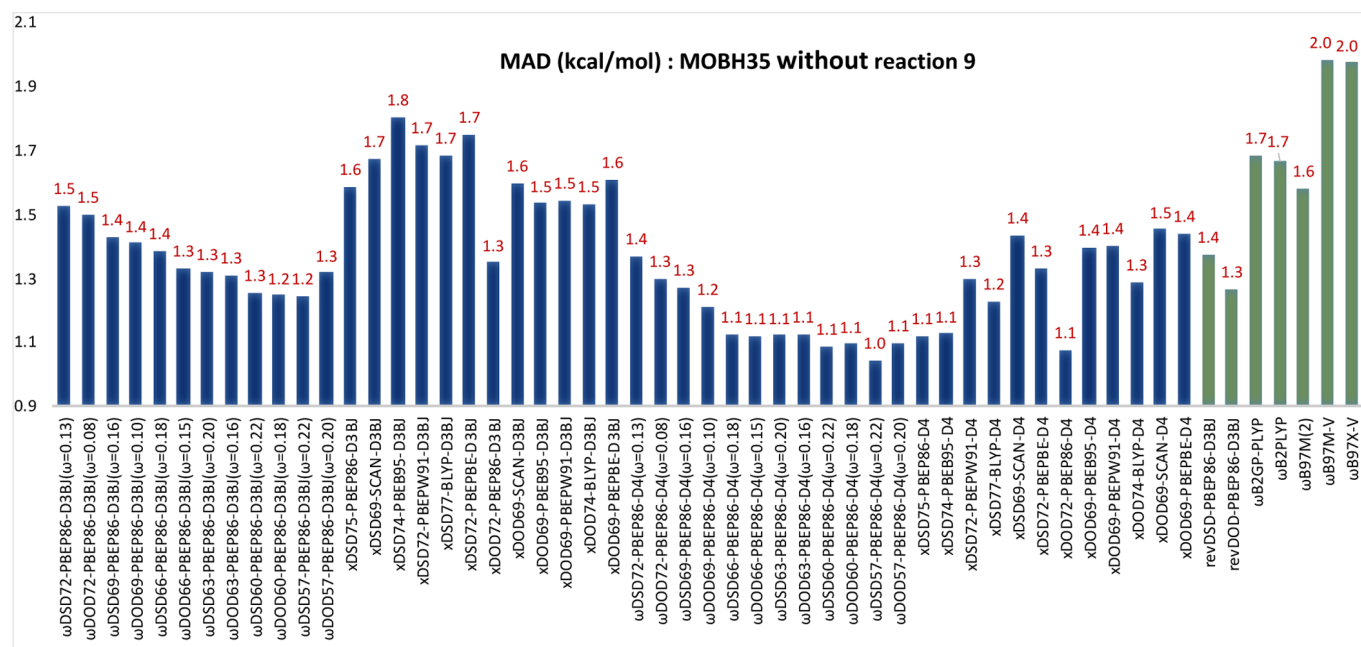

**Figure 2.** MADs (kcal/mol) for our new *x*DSD and *ω*DSD functionals tested against modified MOBH35.

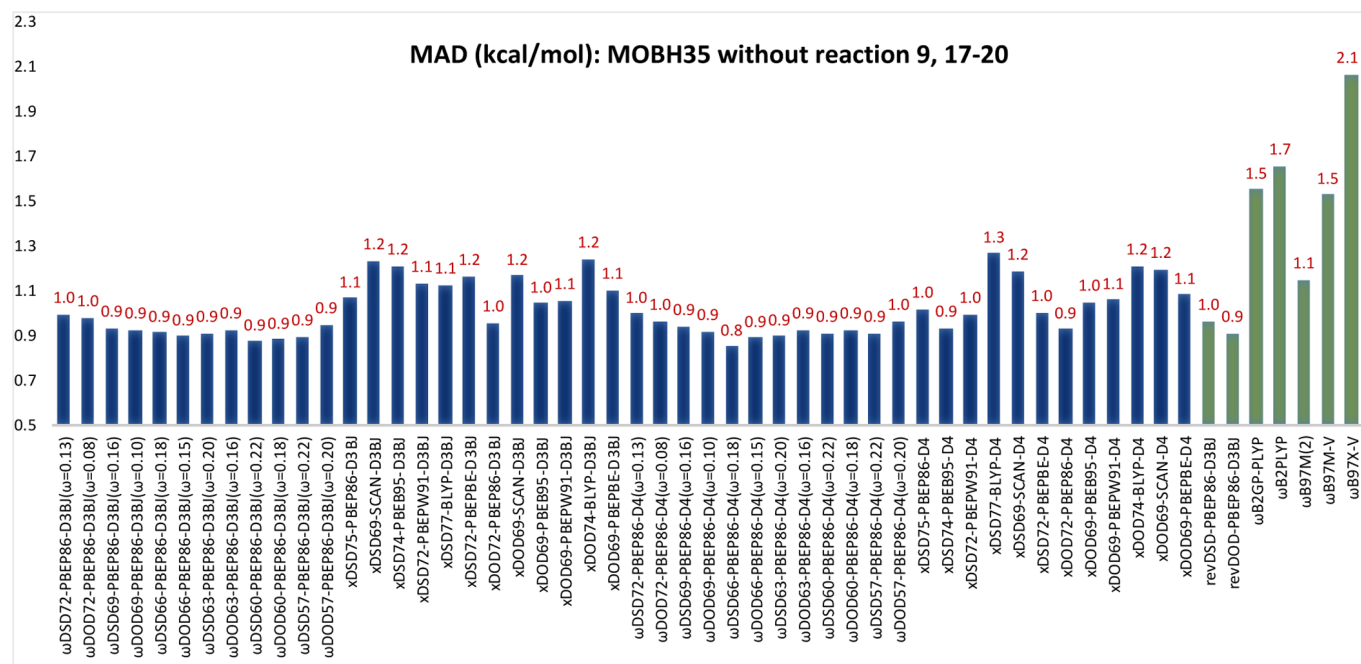

**Figure 3.** MAD (kcal/mol) statistics for *x*DSD and *ω*DSD functionals evaluated against modified MOBH35.

Now, we shift our focus from *ω*DSD-D4 to *ω*DOD-D4 functionals. The *ω*DOD$_{69}$-PBEP86-D4 (*ω* = 0.10) functional is the best performer here with the WTMAD2 = 2.175 kcal/mol. By having one parameter fewer (seven instead of eight), we sacrificed only 0.09 kcal/mol accuracy, and small-molecule thermochemistry is the reason behind this loss (see Table S4).

Similar to *x*DSD cases, here also, including the core−valence correlation for the MP2-like term or considering the MBD term beyond the three-body ATM correction did not help either.

We also considered eliminating the dispersion correction altogether. Similar to the global DH cases,[9] this approach significantly degrades the accuracy here too. The general trend shows the improvement in performance with increased HF exchange and the requirement of a higher *ω* value with respect to their *ω*DSD and *ω*DOD counterparts.

**3.4. Benchmarks External to GMTKN55.** The performances of the newly developed functionals were tested using four datasets, which are external to GMTKN55. These four test sets are MOBH35, originally proposed by Iron and Janes,[88] POLYPYR21, MPCONF196,[89] and CHAL336.[90]

*3.4.1. MOBH35.* This database[88] comprises 35 reactions, including both early and late transition metal groups and 3d, 4d, and 5d transition metals. We extracted the best reported reference energies from the erratum[88] to the original[88]

Table 3. MADs (in kcal/mol) and rmsd Values (in kcal/mol) for the New $x$DSD (DOD) and $\omega$DSD (DOD) Functionals Evaluated against POLYPYR21

| functionals | MAD (kcal/mol) | rmsd (kcal/mol) total | rmsd Möbius structures | rmsd Hückel & figure-eight structures |
|---|---|---|---|---|
| $x$DSD$_{72}$-PBEPBE-D3BJ | 2.31 | 3.33 | 5.36 | 0.90 |
| $x$DOD$_{69}$-PBEPBE-D3BJ | 1.69 | 2.37 | 3.68 | 0.76 |
| $x$DSD$_{74}$-PBEB95-D3BJ | 2.50 | 3.59 | 5.73 | 0.95 |
| $x$DOD$_{69}$-PBEB95-D3BJ | 1.30 | 1.75 | 2.34 | 0.81 |
| $x$DSD$_{72}$-PBEPW91-D3BJ | 2.38 | 3.44 | 5.53 | 0.91 |
| $x$DOD$_{69}$-PBEPW91-D3BJ | 1.63 | 2.27 | 3.47 | 0.75 |
| $x$DSD$_{69}$-SCAN-D3BJ | 1.96 | 2.73 | 4.29 | 0.86 |
| $x$DOD$_{69}$-SCAN-D3BJ | 1.63 | 2.24 | 3.42 | 0.80 |
| $x$DSD$_{77}$-BLYP-D3BJ | 3.08 | 4.51 | 7.28 | 1.14 |
| $x$DOD$_{74}$-BLYP-D3BJ | 1.19 | 1.64 | 2.13 | 0.75 |
| $x$DSD$_{75}$-PBEP86-D3BJ | 2.46 | 3.58 | 5.74 | 0.90 |
| $x$DOD$_{72}$-PBEP86-D3BJ | 1.42 | 1.96 | 2.88 | 0.70 |
| $x$DSD$_{72}$-PBEPBE-D4 | 1.82 | 2.49 | 3.27 | 1.02 |
| $x$DOD$_{69}$-PBEPBE-D4 | 1.72 | 2.39 | 3.55 | 0.79 |
| $x$DSD$_{74}$-PBEB95-D4 | 1.58 | 2.18 | 2.60 | 1.01 |
| $x$DOD$_{69}$-PBEB95-D4 | 1.54 | 2.13 | 2.54 | 1.05 |
| $x$DSD$_{72}$-PBEPW91-D4 | 1.74 | 2.48 | 3.58 | 0.79 |
| $x$DOD$_{69}$-PBEPW91-D4 | 1.68 | 2.35 | 3.43 | 0.80 |
| $x$DSD$_{69}$-SCAN-D4 | 1.76 | 2.42 | 3.25 | 0.97 |
| $x$DOD$_{69}$-SCAN-D4 | 1.63 | 2.28 | 3.27 | 0.82 |
| $x$DSD$_{77}$-BLYP-D4 | 1.56 | 2.20 | 2.49 | 1.01 |
| $x$DOD$_{74}$-BLYP-D4 | 1.31 | 1.87 | 2.23 | 0.86 |
| $x$DSD$_{75}$-PBEP86-D4 | 2.08 | 3.02 | 4.52 | 0.80 |
| $x$DOD$_{72}$-PBEP86-D4 | 1.30 | 1.82 | 2.29 | 0.78 |
| $\omega$DSD$_{72}$-PBEP86-D3BJ ($\omega$ = 0.13) | 1.61 | 2.34 | 3.77 | 0.72 |
| $\omega$DOD$_{72}$-PBEP86-D3BJ ($\omega$ = 0.08) | 1.28 | 1.85 | 3.09 | 0.63 |
| $\omega$DSD$_{69}$-PBEP86-D3BJ ($\omega$ = 0.16) | 0.92 | 1.34 | 2.12 | 0.54 |
| $\omega$DOD$_{69}$-PBEP86-D3BJ ($\omega$ = 0.10) | 0.70 | 1.00 | 1.56 | 0.49 |
| $\omega$DSD$_{66}$-PBEP86-D3BJ ($\omega$ = 0.18) | 0.53 | 0.72 | 1.09 | 0.44 |
| $\omega$DOD$_{66}$-PBEP86-D3BJ ($\omega$ = 0.15) | 0.38 | 0.49 | 0.69 | 0.38 |
| $\omega$DSD$_{63}$-PBEP86-D3BJ ($\omega$ = 0.20) | 0.97 | 1.33 | 1.86 | 0.60 |
| $\omega$DOD$_{63}$-PBEP86-D3BJ ($\omega$ = 0.16) | 0.78 | 1.06 | 1.41 | 0.55 |
| $\omega$DSD$_{60}$-PBEP86-D3BJ ($\omega$ = 0.22) | 0.38 | 0.49 | 0.45 | 0.44 |
| $\omega$DOD$_{60}$-PBEP86-D3BJ ($\omega$ = 0.18) | **0.35** | **0.45** | 0.49 | 0.41 |
| $\omega$DSD$_{57}$-PBEP86-D3BJ ($\omega$ = 0.22) | 0.44 | 0.60 | 0.95 | 0.38 |
| $\omega$DOD$_{57}$-PBEP86-D3BJ ($\omega$ = 0.20) | 0.55 | 0.75 | 1.26 | 0.36 |
| $\omega$DSD$_{72}$-PBEP86-D4 ($\omega$ = 0.13) | 1.49 | 2.16 | 3.32 | 0.59 |
| $\omega$DOD$_{72}$-PBEP86-D4 ($\omega$ = 0.08) | 1.08 | 1.58 | 2.36 | 0.50 |
| $\omega$DSD$_{69}$-PBEP86-D4 ($\omega$ = 0.16) | 0.84 | 1.25 | 1.80 | 0.46 |
| $\omega$DOD$_{69}$-PBEP86-D4 ($\omega$ = 0.10) | 0.56 | 0.84 | 1.08 | 0.43 |
| $\omega$DSD$_{66}$-PBEP86-D4 ($\omega$ = 0.18) | 0.43 | 0.59 | 0.67 | 0.40 |
| $\omega$DOD$_{66}$-PBEP86-D4 ($\omega$ = 0.15) | 0.42 | 0.55 | 0.64 | 0.42 |
| $\omega$DSD$_{63}$-PBEP86-D4 ($\omega$ = 0.20) | 0.99 | 1.39 | 1.70 | 0.64 |
| $\omega$DOD$_{63}$-PBEP86-D4 ($\omega$ = 0.16) | 0.82 | 1.16 | 1.34 | 0.59 |
| $\omega$DSD$_{60}$-PBEP86-D4 ($\omega$ = 0.22) | 0.45 | 0.65 | 0.41 | 0.50 |
| $\omega$DOD$_{60}$-PBEP86-D4 ($\omega$ = 0.18) | 0.43 | 0.59 | 0.47 | 0.48 |
| $\omega$DSD$_{57}$-PBEP86-D4 ($\omega$ = 0.22) | 0.55 | 0.55 | 0.99 | 0.48 |
| $\omega$DOD$_{57}$-PBEP86-D4 ($\omega$ = 0.20) | 0.63 | 0.83 | 1.32 | 0.48 |
| $\omega$B97M(2)[93] | 0.48 | 0.63 | 0.82 | 0.55 |
| $\omega$B2PLYP[93] | 0.97 | 1.28 | 2.10 | 0.62 |
| $\omega$B2GP-PLYP[93] | 0.61 | 0.78 | 0.99 | 0.57 |

MOBH35 article. The def2-QZVPP basis set was used with grids and auxiliary basis sets as described above in Section 2.2.

Using a variety of multireference diagnostics, our group has recently found (E. Semidalas and J.M.L. Martin, unpublished) that reaction 9 exhibits severe static correlation in all three structures, which gets progressively worse from the reactant via the transition state to the product as the HOMO−LUMO gap narrows. Under these circumstances, as previously found for polypyrroles,[91] a large gap opens between canonical CCSD(T) and DLPNO-CCSD(T), and yet for the product, diagnostics are so large that one can legitimately question whether CCSD(T) itself is adequate for the problem. Hence, omitting this reaction from the MOBH35 dataset, we have recalculated MAD values for the remaining 34 reactions (see Figure 2).

In general, with the D3BJ dispersion correction, both range-separated and global DOD functionals perform better than their DSD counterparts. Shifting from D3BJ to D4 benefits the $\omega$DSD ($\omega$DOD) functionals across the board by 0.2−0.3 kcal/mol. $\omega$DSD$_{57}$-PBEP86-D4 ($\omega$ = 0.22) achieves the lowest MAD of 1.0 kcal/mol, closely followed by the other range-separated DSD (DOD) functionals. Therefore, there is very little to choose among them.

Among the $x$DSD ($x$DOD) functionals with D3BJ, $x$DODs still do better than the $x$DSD variants. However, when we substitute the D4 dispersion correction, $x$DSDs are better performers than $x$DODs. The only exception is the PBEP86 XC combination, where $x$DOD$_{72}$-PBEP86-D4 marginally outperforms $x$DSD$_{75}$-PBEP86-D4 (see Figure 2). $x$DOD$_{72}$-PBEP86-D4 offers the lowest MAD of 1.1 kcal/mol, close to what was found for $\omega$DSD$_{57}$-PBEP86-D4 ($\omega$ = 0.22). That being said, the other empirical range-separated double hybrids $\omega$B2PLYP, $\omega$B2GP-PLYP, and $\omega$B97M(2) all have larger MAD values in the 1.6−1.7 kcal/mol range (see Figure 2). Finally, we can conclude that for the PBEP86 XC combination, shifting from the global to range-separated double hybrid is not so beneficial. Similar to what we found for the GMTKN55 test suite, considering higher-order MBD terms beyond the three-body ATM term has no perceptible benefit, though again, the systems under investigation may simply be too small.

Now, the bimolecular reactions (i.e., reaction 17−20) could be problematic for a different reason, that is, because of proneness to basis set superposition error (BSSE). (We note that these reactions were omitted from Dohm et al.'s recent revision[92] of MOBH35.) Therefore, if we also drop reactions 17–20 together with reaction 9 and recalculate MADs for the remaining 30 reactions, the MAD drops across the board. Also, while the MAD for ωB2PLYP and ωB2GP-PLYP remains elevated, that for ωB97M(2) now is in the same cohort as those of our best functionals (see Figure 3).

Table 4. MADs (in kcal/mol) and rmsds (in kcal/mol) of Conformational Energies for Various DHDFs Evaluated against the MPCONF196 Dataset

| Functionals | MAD (kcal/Mol) | RMSD(kcal/mol) | | | | Functionals | MAD (kcal/mol) | RMSD (kcal/mol) | | | |
|---|---|---|---|---|---|---|---|---|---|---|---|
| | | Total | Small[a] | Medium[b] | Large[c] | | | Total | Small[a] | Medium[b] | Large[c] |
| ωB97X-V | 0.45 | 0.73 | 0.19 | 0.50 | 1.36 | xDOD$_{69}$-PBEPBE-D3BJ | 0.34 | 0.53 | 0.20 | 0.33 | 0.97 |
| ωB97M-V | 0.55 | 0.87 | 0.27 | 0.61 | 1.60 | xDOD$_{69}$-PBEPBE-D4 | 0.28 | 0.42 | 0.18 | 0.32 | 0.73 |
| ωB97M(2) | 0.37 | 0.62 | 0.16 | 0.38 | 1.17 | xDSD$_{74}$-PBEB95-D3BJ | 0.46 | 0.66 | 0.35 | 0.46 | 1.16 |
| revDSD-PBEP86-D3BJ | 0.29 | 0.46 | 0.17 | 0.25 | 0.89 | xDSD$_{74}$-PBEB95-D4 | 0.34 | 0.45 | 0.31 | 0.41 | 0.67 |
| revDSD-PBEP86-D4 | 0.26 | 0.38 | 0.17 | 0.25 | 0.69 | xDOD$_{69}$-PBEB95-D3BJ | 0.49 | 0.69 | 0.38 | 0.51 | 1.19 |
| revDOD-PBEP86-D3BJ | 0.29 | 0.47 | 0.17 | 0.26 | 0.90 | xDOD$_{69}$-PBEB95-D4 | 0.38 | 0.54 | 0.29 | 0.45 | 0.88 |
| revDOD-PBEP86-D4 | 0.26 | 0.38 | 0.17 | 0.26 | 0.67 | xDSD$_{72}$-PBEPW91-D3BJ | 0.33 | 0.52 | 0.19 | 0.31 | 0.99 |
| revDSD-PBEPBE-D3BJ | 0.34 | 0.53 | 0.22 | 0.43 | 0.90 | xDSD$_{72}$-PBEPW91-D4 | 0.26 | 0.37 | 0.19 | 0.31 | 0.62 |
| revDSD-PBEPBE-D4 | 0.29 | 0.44 | 0.17 | 0.41 | 0.72 | xDOD$_{69}$-PBEPW91-D3BJ | 0.33 | 0.52 | 0.20 | 0.32 | 0.97 |
| revDOD-PBEPBE-D3BJ | 0.34 | 0.53 | 0.22 | 0.44 | 0.90 | xDOD$_{69}$-PBEPW91-D4 | 0.28 | 0.41 | 0.18 | 0.31 | 0.73 |
| revDOD-PBEPBE-D4 | 0.29 | 0.44 | 0.17 | 0.41 | 0.72 | ωDSD$_{72}$-PBEP86-D3BJ(ω=0.13) | 0.32 | 0.53 | 0.17 | 0.29 | 1.01 |
| DOD66-SCAN-D3BJ | 0.32 | 0.49 | 0.21 | 0.29 | 0.92 | ωDSD$_{72}$-PBEP86-D4(ω=0.13) | 0.27 | 0.40 | 0.18 | 0.27 | 0.73 |
| DOD66-SCAN-D4 | 0.27 | 0.38 | 0.20 | 0.30 | 0.64 | ωDOD$_{72}$-PBEP86-D3BJ(ω=0.08) | 0.32 | 0.53 | 0.17 | 0.30 | 1.00 |
| revDSD-PBEB95-D3BJ | 0.44 | 0.61 | 0.35 | 0.45 | 1.05 | ωDOD$_{72}$-PBEP86-D4(ω=0.08) | 0.27 | 0.40 | 0.18 | 0.27 | 0.71 |
| revDSD-PBEB95-D4 | 0.36 | 0.50 | 0.29 | 0.43 | 0.80 | ωDSD$_{69}$-PBEP86-D3BJ(ω=0.16) | 0.32 | 0.53 | 0.17 | 0.29 | 1.01 |
| revDOD-PBEB95-D3BJ | 0.46 | 0.64 | 0.37 | 0.48 | 1.08 | ωDSD$_{69}$-PBEP86-D4(ω=0.16) | 0.27 | 0.40 | 0.17 | 0.27 | 0.72 |
| revDOD-PBEB95-D4 | 0.36 | 0.51 | 0.29 | 0.44 | 0.81 | ωDOD$_{69}$-PBEP86-D3BJ(ω=0.10) | 0.32 | 0.52 | 0.17 | 0.30 | 1.00 |
| revDSD-BLYP-D3BJ | 0.35 | 0.58 | 0.16 | 0.31 | 1.11 | ωDOD$_{69}$-PBEP86-D4(ω=0.10) | 0.27 | 0.39 | 0.18 | 0.27 | 0.71 |
| revDSD-BLYP-D4 | 0.29 | 0.44 | 0.18 | 0.29 | 0.81 | ωDSD$_{66}$-PBEP86-D3BJ(ω=0.18) | 0.32 | 0.53 | 0.17 | 0.29 | 1.01 |
| revDOD-BLYP-D3BJ | 0.36 | 0.60 | 0.16 | 0.34 | 1.15 | ωDSD$_{66}$-PBEP86-D4(ω=0.18) | 0.27 | 0.41 | 0.17 | 0.27 | 0.74 |
| revDOD-BLYP-D4 | 0.30 | 0.46 | 0.20 | 0.31 | 0.83 | ωDOD$_{66}$-PBEP86-D3BJ(ω=0.15) | 0.33 | 0.53 | 0.17 | 0.31 | 1.01 |
| revDSD-PBEPW91-D3BJ | 0.33 | 0.50 | 0.21 | 0.32 | 0.92 | ωDOD$_{66}$-PBEP86-D4(ω=0.15) | 0.27 | 0.40 | 0.17 | 0.28 | 0.72 |
| revDSD-PBEPW91-D4 | 0.27 | 0.40 | 0.18 | 0.31 | 0.70 | ωDSD$_{63}$-PBEP86-D3BJ(ω=0.20) | 0.32 | 0.53 | 0.17 | 0.30 | 1.01 |
| revDOD-PBEPW91-D3BJ | 0.32 | 0.50 | 0.21 | 0.32 | 0.91 | ωDSD$_{63}$-PBEP86-D4(ω=0.20) | 0.27 | 0.41 | 0.17 | 0.28 | 0.74 |
| revDOD-PBEPW91-D4 | 0.28 | 0.40 | 0.17 | 0.32 | 0.70 | ωDOD$_{63}$-PBEP86-D3BJ(ω=0.16) | 0.33 | 0.53 | 0.17 | 0.31 | 1.01 |
| xDSD$_{75}$-PBEP86-D3BJ | 0.31 | 0.50 | 0.16 | 0.27 | 0.96 | ωDOD$_{63}$-PBEP86-D4(ω=0.16) | 0.27 | 0.41 | 0.17 | 0.28 | 0.74 |
| xDSD$_{75}$-PBEP86-D4 | 0.25 | 0.35 | 0.20 | 0.27 | 0.59 | ωDSD$_{60}$-PBEP86-D3BJ(ω=0.22) | 0.32 | 0.53 | 0.17 | 0.30 | 1.01 |
| xDOD$_{72}$-PBEP86-D3BJ | 0.31 | 0.51 | 0.16 | 0.29 | 0.97 | ωDSD$_{60}$-PBEP86-D4(ω=0.22) | 0.27 | 0.40 | 0.17 | 0.27 | 0.74 |
| xDOD$_{72}$-PBEP86-D4 | 0.25 | 0.36 | 0.19 | 0.27 | 0.62 | ωDOD$_{60}$-PBEP86-D3BJ(ω=0.18) | 0.32 | 0.53 | 0.17 | 0.31 | 1.00 |
| xDSD$_{69}$-SCAN-D3BJ | 0.32 | 0.51 | 0.20 | 0.30 | 0.96 | ωDOD$_{60}$-PBEP86-D4(ω=0.18) | 0.27 | 0.41 | 0.17 | 0.29 | 0.74 |
| xDOD$_{69}$-SCAN-D3BJ | 0.34 | 0.53 | 0.21 | 0.31 | 0.99 | ωDSD$_{57}$-PBEP86-D3BJ(ω=0.22) | 0.35 | 0.57 | 0.19 | 0.35 | 1.08 |
| xDSD$_{69}$-SCAN-D4 | 0.26 | 0.37 | 0.20 | 0.30 | 0.62 | ωDSD$_{57}$-PBEP86-D4(ω=0.22) | 0.27 | 0.43 | 0.16 | 0.30 | 0.79 |
| xDOD$_{69}$-SCAN-D4 | 0.27 | 0.39 | 0.20 | 0.29 | 0.67 | ωDOD$_{57}$-PBEP86-D3BJ(ω=0.20) | 0.33 | 0.54 | 0.18 | 0.32 | 1.02 |
| xDSD$_{77}$-BLYP-D3BJ | 0.36 | 0.59 | 0.18 | 0.32 | 1.13 | ωDOD$_{57}$-PBEP86-D4(ω=0.20) | 0.27 | 0.41 | 0.17 | 0.29 | 0.73 |
| xDSD$_{77}$-BLYP-D4 | 0.32 | 0.40 | 0.32 | 0.42 | 0.49 | ωB2PLYP | 0.55 | 0.78 | 0.40 | 0.71 | 1.25 |
| xDOD$_{74}$-BLYP-D3BJ | 0.36 | 0.60 | 0.17 | 0.35 | 1.14 | ωB2GP-PLYP | 0.49 | 0.74 | 0.27 | 0.65 | 1.25 |
| xDOD$_{74}$-BLYP-D4 | 0.31 | 0.45 | 0.21 | 0.31 | 0.82 | PBE0-D3(0)[d] | 0.50 | 0.64 | 0.50 | 0.50 | 0.93 |
| xDSD$_{72}$-PBEPBE-D3BJ | 0.34 | 0.53 | 0.20 | 0.32 | 0.99 | PBE-D3(0)[d] | 0.69 | 0.83 | 0.78 | 0.70 | 1.04 |
| xDSD$_{72}$-PBEPBE-D4 | 0.27 | 0.38 | 0.19 | 0.32 | 0.64 | | | | | | |

[a]Small subsets: FGG, GGF, WG, WGG, and GFA. [b]Medium subsets: POXTRD, CAMVES, COHVAW, CHPSAR, and Cpd_B. [c]Large subsets: Cpd_A, SANGLI, and YIVNOG. [d]MAD and rmsd values are taken from ref 89.

*3.4.2. POLYPYR21.* This database[91] contains 21 unique structures of penta-, hexa- and heptaphyrins, which are [4n] π-electron expanded porphyrins that have generated considerable interest recently because of their potential application as molecular switches (see the introduction to ref 93 for a brief review). The structures are Hückel, Möbius, and figure-eight minima as well as the various transition states between them. Among them, the most troublesome are the Möbius rings, which exhibit a pronounced multireference character (for more details, see refs[91,93]).

CCSD(T)/CBS level reference energies were extracted from ref 93. We have used the def2-TZVP basis for all calculations here.

With the D3BJ dispersion correction, it appears that xDOD functionals perform noticeably better than their xDSD counterparts. In ref 93 for the problem at hand, as well as in the study by Iron and Janes[88] for MOBH35, the same trend was observed for DOD versus DSD functionals and ascribed to the greater resilience of spin-opposite-scaled GLPT2 to the static correlation. Now, if here we replace D3BJ by D4, the large difference between rmsd values of xDSD and xDOD functionals goes away (see Table 3). Finally, judging from the rmsd error statistics listed in Table 3, we observe that xDOD$_{74}$-BLYP-D3BJ offers the lowest rmsd (1.64 kcal/mol) among the xDSD functionals.

In general, range-separated DSD double hybrids are better performers than the xDSD or xDOD variants (see Table 3). Switching from D3BJ to D4 dispersion correction deteriorates the performance for the ωDOD functional variants. Among all the xDSD (xDOD) and ωDSD (ωDOD) functionals tested, ωDOD$_{63}$-PBEP86-D3BJ (ω = 0.16) offers the lowest rmsd = 0.45 kcal/mol, which, in fact, slightly outperforms the previously reported[93] top performer ωB97M(2) (0.63 kcal/mol). However, in light of remaining uncertainties in the reference values, this difference should not be considered significant. Inspection of the Möbius structure data in isolation does reveal, across the board, that range-separated DHs cope with them much better than global double hybrids.

*3.4.3. MPCONF196.* This database[89] contains a set of carefully selected five di- and tripeptides (namely, FGG, GGF, WG, WGG, and GFA, see ref 94 for more details) and eight macrocycles, comprising 13 compounds in total. Among the eight macrocycles, five compounds with different ring sizes (Cambridge structural database[95] acronyms: POXTRD, CAMVES, COHVAW, CHPSAR, and YIVNOG) are taken from ref 96; the next two compounds are inhibitors of human cyclophilin A: sanglifehrin A analogue (SANGLI)[97] and Cpd_A,[98] while the final compound is the acyclic synthetic precursor of Cpd_A, denoted Cpd_B (for more details, see Figure 1 of ref 89). Considering both high- and low-energy conformers (15 or 16 for each compound) MPCONF196 consists of a total of 196 unique structures.

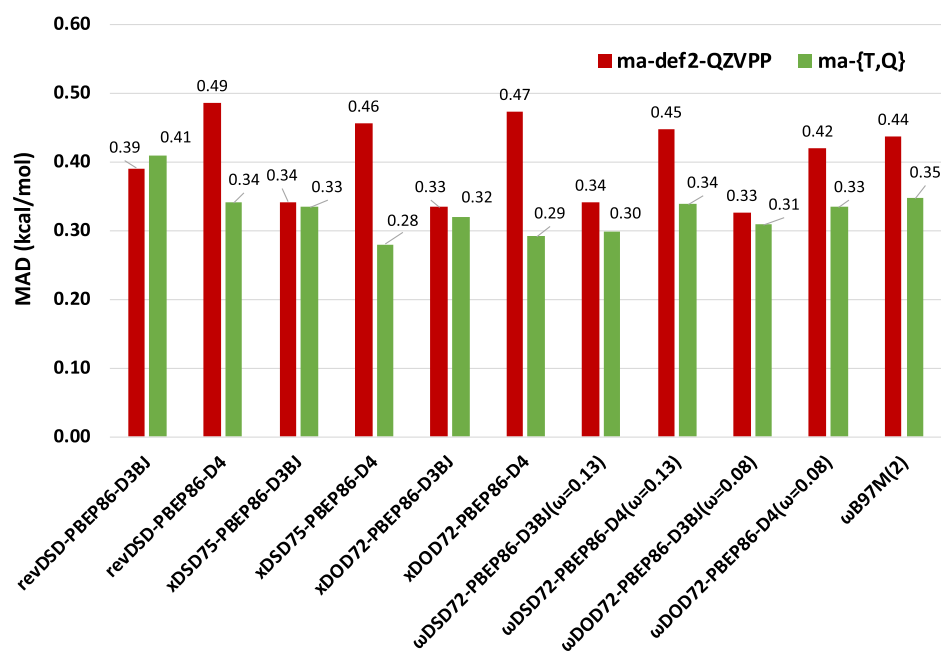

**Figure 4.** MAD (kcal/mol) statistics of the selected global and range-separated DSD (DOD) functionals for the CHAL336 benchmark set.

The CCSD(T)/CBS level reference energies for the small systems and their DLPNO-CCSD(T)/CBS[99] counterparts for the larger systems (60−120 atoms) were extracted from the Supporting Information of ref 89. Together with all the new DHDFs we propose in the current study, we also report here the error statistics for our revDSD (revDOD) double hybrids,[9] Mardirossian and Head-Gordon's $\omega$B97M(2),[53] $\omega$B97M-V,[50] and $\omega$B97X-V,[49] and Lars Goerigk and coworkers' $\omega$B2GP-PLYP[58] and $\omega$B2PLYP[58] functionals. All single-point energies were calculated using the def2-TZVPP[69] basis set throughout.

With the D3BJ dispersion correction, all new $\omega$DSD and $\omega$DOD functionals offer almost identical performance. Unlike what we saw for previous two external datasets, performance-wise, there is practically no difference between revDSD and revDOD variants of a specific XC combination.

Now, considering D4 instead of D3BJ benefits both range-separated and global DSD double hybrids across the board. As expected, small subsets are the least and large subsets are the most benefitted cases by this change. Specifically for large subsets, $x$DSD77-BLYP-D4 offers the lowest rmsd (0.49 kcal/mol). However, for small subsets, the PBEB95 XC combination shows a particularly poor performance, even worse than that of lower-rung $\omega$B97X-V. Between $\omega$B97X-V and $\omega$B97M-V, the former functional offers lower rmsd than the latter. Our new $\omega$DSD ($\omega$DOD)-PBEP86-D4 functionals are better performers than the combinatorially optimized, range-separated double hybrid, $\omega$B97M(2).

For the PBEP86 XC combination, shifting from a global to a range-separated DSD-type functional does not offer any improvement in rmsd statistics. Finally, $x$DSD75-PBEP86-D4 is our best pick for the MPCONF196 database. However, we must acknowledge that most of the other global and range-separated DSD-D4 functionals are close contenders (see Table 4).

Similar to the previous two external datasets, and presumably for the same reasons, no benefit is seen from going beyond the three-body ATM term in D4.

**3.4.4. CHAL336.** While the present article was in peer review, Mehta et al.[90] proposed a comprehensive database of chalcogen bonding interactions, CHAL336. Consisting of molecules up to 49 atoms, CHAL336 contains 336 dimers, and a complete evaluation requires 1008 single-point energy calculations. These 336 dimer interaction energies can be subdivided into four categories: chalcogen−chalcogen, chalcogen−$\pi$, chalcogen−halogen, and chalcogen−nitrogen interactions. Mehta et al. have already assessed the performance of a large number of DFT methods (see Figure 9 of their article[90]), and our revDSD-PBEP86-D3BJ[9] was among the best three performers (the differences between these are arguably a photo finish). Here, we evaluate the performance for CHAL336 of eight selected functionals, namely, $x$DSD$_{75}$-PBEP86-D3BJ (D4), $x$DOD$_{72}$-PBEP86-D3BJ (D4), $\omega$DSD$_{72}$-PBEP86-D3BJ (D4), and $\omega$DOD$_{72}$-PBEP86-D3BJ (D4). While we were at it, we also tested revDSD-PBEP86-D3BJ and revDSD-PBEP86-D4, the former to check the consistency with ref 90 and the latter as the D4 variant had not been included in ref 90.

For the $x$DSD ($x$DOD) and $\omega$DSD ($\omega$DOD) functionals, we have used QCHEM 5.3,[67] whereas for the revDSD functional, we have used ORCA4 with the RIJCOSX (chain of spheres[100]) approximation with the most accurate GRIDXS9. The same minimally augmented diffuse def2 basis set as in ref 90, ma-def2-QZVPP,[101] has been used across the board.

We found both $x$DSD75-PBEP86-D3BJ and $x$DOD72-PBEP86-D3BJ to perform slightly better than revDSD-PBEP86-D3BJ (see Figure 4 and Table S17 in the Supporting Information). However, for the subsets of systems where both canonical and DLPNO reference values were available in ref 90, the rmsd between them is 0.15 kcal/mol, which leads us to assume an uncertainty of about 0.15 kcal/mol in the reference values (see Tables 1 and S4 in ref 90). Hence, the apparent improvement in the statistics of $x$DSD ($x$DOD) and $\omega$DSD ($\omega$DOD) functionals arguably does not rise above numerical noise.

From the ma-def2-QZVPP and ma-def2-TZVPP energies, two-point CBS extrapolation has been performed for each of the above functionals using the $L^{-3}$ formula (where $L$ is the cardinal number of the basis set), which works out in practice to $E\infty \approx E[Q] + A(E[Q] - E[T])$, where $A = ((4/3)^3 - 1)^{-1} = 0.7297$. CBS extrapolation significantly improved the performance for both $x$DSD75-PBEP86-D4 (the MAD improves from 0.46 to 0.28 kcal/mol) and $x$DOD72-PBEP86-D4 (the MAD drops from 0.47 to 0.29 kcal/mol). Suffice to say that all functionals considered here are at least competitive with, and perhaps superior to, the best performers in the CHAL336 article, despite this dataset not having been involved in parameterization.

## 4. CONCLUSIONS

Aiming to improve our previous revDSD family functionals further, we have considered both range separation $\omega$DSDs and $x$DSDs—the latter, while analogues of the XYG3 family of functionals, are also recovered as the $\omega = 0$ limit of range-separated double hybrids. Concerning our first research objective: from an extensive survey, we can conclude the following:

(a) $x$DSD-D3BJ functionals have a slight advantage over our prior revDSD family functionals, which can be further improved upon by replacing D3BJ with D4.

(b) For D4, allowing $c_{ATM}$, the prefactor for the three-body ATM term, to take on values different from one does not reduce WTMAD2 by an amount statistically significant enough that it would justify the introduction of the extra adjustable parameter. Replacing the ATM term by the many-body dispersion model of Tkatchenko and coworkers achieves no significant benefit, although the systems in GMTKN55 may simply not be large enough to rule this out.

(c) For the $x$DSD$_n$-PBEP86-D4 variants, $c_{X,HF} = 0.75$ offers the lowest WTMAD2, unlike the previously reported $c_{X,HF} = 0.69$ for DSD-PBEP86-D4. However, when we imposed the $c_{2ss} = 0$ constraint, WTMAD2 reaches a minimum at $c_{X,HF} = 0.72$.

(d) In terms of WTMAD2, $x$DSD$_{75}$-PBEP86-D4 marginally outperforms $\omega$B97M(2),[53] hitherto the "record holder" for the lowest WTMAD2,[8] but without its range separation and using just a half-dozen empirical parameters. In view of the uncertainty in the reference values, however, and the fact that $x$DSD$_{75}$-PBEP86-D4 was trained on GMTKN55 itself rather than on a different albeit strongly overlapping set such as $\omega$B97M(2), it is probably safer to say that the two functionals are competitive.

Concerning the second research objective, applying range separation over the HF exchange part, we found the lowest WTMAD2 for $c_{X,HF} = 0.72$ and $\omega = 0.13$. With D3BJ, WTMAD2 is 2.108 kcal/mol, which can be lowered slightly further by substituting D4 (2.083 kcal/mol). Therefore, range separation helped us to improve the performance slightly beyond that of $x$DSD$_{75}$-PBEP86-D3BJ(D4) and in turn a little further beyond that of $\omega$B97M(2)—again using just a half-dozen adjustable parameters. Although $\omega$B97M(2) outperforms all the new $\omega$DSD and $\omega$DSD functionals for small-molecule thermochemistry, this is outweighed in WTMAD2 by the superior performance of the new functionals for conformer equilibria.

All in all, however, the improvement for GMTKN55 from introducing range separation in DSD functionals is quite modest, somewhat surprisingly so. In some sense, this is convenient as GHs tend to be computationally more economical.

For some perspective beyond the comparison of WTMAD2 [where differences between, for example, $\omega$B97M(2) and $\omega$DSD$_{72}$-PBEP86-D4 ($\omega = 0.13$) may well be comparable to the residual uncertainty in the reference values], let us consider the performance of four representative functionals, namely, $\omega$B97M(2), revDSD-PBEP86-D4, $x$DSD$_{75}$-PBEP86-D4, and $\omega$DSD$_{72}$-PBEP86-D4 ($\omega = 0.13$), for four external test sets not involved in the parameterization process. Two of these, metal–organic barrier heights (MOBH35)[88] and especially the isomer equilibria and interconversion barriers in polypyrroles (POLYPYR21),[91,93] put the functionals' performance in the presence of static correlation to the test. The two others are CHAL336, a very recently published[90] benchmark of chalcogen bonding interactions, and MPCONF196,[89] which features conformational energies of smaller peptides and of medium-sized macrocycles.

All four options perform very well for MOBH35 if pathologically multireference reaction 9 and bimolecular reactions 17−20 are removed (see above). For CHAL336, again, all four options perform excellently if the basis set extrapolation is performed to quench the effect of the BSSE; two of the options have smaller MAD values by about 0.05 kcal/mol, but in light of the residual uncertainty of about 0.15 kcal/mol rmsd in the reference data, this difference may be deemed insignificant.

For MPCONF196, $\omega$B97M(2) still performs very well but less so than the other three options. This is consistent, actually, with the breakdown of WTMAD2 for GMTKN55 into five top-level subcategories: $\omega$B97M(2) has an edge there for the small-molecule thermochemistry component, which is compensated by the better performance for the intramolecular interaction component.

This leaves POLYPYR21, where the two range-separated options are clearly superior and pronouncedly so for the Möbius structures (which have very pronounced static correlation[91,93]). It may therefore be that range-separated double hybrids, be they $\omega$B97M(2) or $\omega$DSD, have an edge for these kinds of problems; moreover, as will be seen in the companion article (part II), the combination of range separation with post-PT2 corrections turns out to be more generally advantageous.[20]

## ■ ASSOCIATED CONTENT

### ⓈⒾ Supporting Information

The Supporting Information is available free of charge at https://pubs.acs.org/doi/10.1021/acs.jpca.1c01294.

Abridged details of all 55 subsets of GMTKN55 with proper references; the benefit of using $c_{ATM}$ and $c_{MBD}$ for the revDSD functionals; prescreening for wDSD$_n$-PBEB95-D3BJ, wDSD$_n$-PBEPW91-D3BJ, wDSD$_n$-PBEPBE-D3BJ, and wDOD$_n$ variants using diet100;

breakdown of total WTMAD2 into five top-level subsets for all the *x*DSD and range-separated DSD functionals; MAD, MSD, and rmsd values as well as breakdown of total WTMAD2 by each subset for new functionals; MAD statistics of various DSD-DHs for the full CHAL336 and its four subsets; and QCHEM sample inputs for *ω*DSD (*ω*DOD)-D3BJ and ORCA sample inputs for *x*DSD (*x*DOD)-D3BJ functionals (PDF)

■ AUTHOR INFORMATION

**Corresponding Author**

Jan M. L. Martin − *Department of Organic Chemistry, Weizmann Institute of Science, 7610001 Reḥovot, Israel;* orcid.org/0000-0002-0005-5074; Email: gershom@weizmann.ac.il

**Authors**

Golokesh Santra − *Department of Organic Chemistry, Weizmann Institute of Science, 7610001 Reḥovot, Israel;* orcid.org/0000-0002-7297-8767

Minsik Cho − *Department of Organic Chemistry, Weizmann Institute of Science, 7610001 Reḥovot, Israel; Department of Chemistry, Brown University, Providence, Rhode Island 02912, United States;* orcid.org/0000-0002-9307-8549

Complete contact information is available at:
https://pubs.acs.org/10.1021/acs.jpca.1c01294

**Funding**

This research was funded by the Israel Science Foundation (grant 1969/20) and by the Minerva Foundation (grant 20/05).

**Notes**

The authors declare no competing financial interest.

■ ACKNOWLEDGMENTS

We would like to acknowledge helpful discussions with Emmanouil Semidalas (WIS), Dr. Mark A. Iron (WIS), and Prof. Mercedes Alonso Giner (the Free University of Brussels). We also thank Nitai Sylvetsky for critical comments on the manuscript draft. G.S. acknowledges a doctoral fellowship from the Feinberg Graduate School (WIS). M.C. was a Karyn Kupcynet-Getz International Summer School Fellow at WIS in 2019.